# Speed Control of Permanent Magnet DC Motor with friction and measurement noise using Novel Nonlinear Extended State Observer Based Anti-Disturbance control

**Amjad J. Humaidi** [1], **Ibraheem Kasim Ibraheem** [2*]

[1] Department of Control and Systems Engineering, The University of Technology, Baghdad 10001, Iraq; 601116@uotechnology.edu.iq

[2] The University of Baghdad, College of Engineering, Department of Electrical Engineering, Al-Jadriyah, P.O.B.: 47273, 10001 Baghdad, Iraq.

* Correspondence: ibraheemki@coeng.uobaghdad.edu.iq; ( 00964 7702562658)

**Abstract:** In this paper, a novel finite-time Nonlinear Extended State Observer (NLESO) is proposed and employed in Active Disturbance Rejection Control (ADRC) to stabilize a nonlinear system against system's uncertainties and discontinuous disturbances using output feedback based control. The first task was to aggregate the uncertainties, disturbances, and any other undesired nonlinearities in the system into a single term called the "generalized disturbance". Consequently, the ESO estimates the generalized disturbance and cancel it from the input channel in an online fashion. A peaking phenomenon that existed in Linear ESO (LESO) has been reduced significantly by adopting a saturation-like nonlinear function in the proposed Nonlinear ESO (NLESO). Stability analysis of the NLEO is studied using finite-time Lyapunov theory, and the comparisons are presented over simulations on Permanent Magnet DC (PMDC) motor to confirm the effectiveness of the proposed observer concerning LESO.

---

## 1. Introduction

In most control industries, it is hard to establish accurate mathematical models to describe the systems precisely. Also, some terms are not explicitly known in mathematical equations and, on the other hand, some unknown external disturbances exist around the system environment. The uncertainty, which includes internal uncertainty and external disturbance, is ubiquitous in practical control systems [1].

To tackle the shortcomings of Passive Anti-Disturbance Control (PADC) strategies in treating the disturbances, Han in [2] has proposed the alleged Active Disturbance Rejection Control (ADRC) paradigm. Generally speaking, the basic idea behind the ADRC is to straightforwardly oppose disturbances by feedforward compensation control configuration given disturbance estimations /cancellation principle [3]. This procedure has discovered an expansive number of practical applications [4,5]. Extended State Observer (ESO) is the central part of the ADRC, which works by augmenting the mathematical model of the nonlinear dynamical system with an additional virtual state. It describes all the unwanted dynamics, uncertainties, and exogenous disturbances and is called the "*generalized disturbance*" or "*total disturbance*". Then this total disturbance is estimated by the ESO and feedback into the control input channel for cancelation [2,6].

During recent years, the literature has found many types of research for analysis, design, and implementation of the ESO. Different forms of modern control theory based observers have been proposed to meet this need where several surveys of various disturbance observers can be found in [7–9]. In 1971, Johnson introduced the Unknown \ Input Observer (UIO) [10] to estimate the unknown input of the system. The transfer function based Disturbance Observer (DOB) [11], proposed later on by Japanese researchers, can predict the disturbance as well. The Perturbation observer (POB) was offered by Kwon and Chung in 2002 [12] in the discrete form to estimate the perturbation acted on the system. The "unknown input", the "disturbance" and the "perturbation" are just different names for the external disturbance, and the above observers can only deal with it. The aforementioned DOBC techniques employ some plant information for disturbance observation and control design. The ESO is the one that uses less information as only the system relative degree should be known (defined later in this paper). For that reason, the ESO has become very popular in recent years. ADRC can be understood as a combination of an extended state observer (ESO) and a state feedback controller, where the ESO is utilized to observe the generalized disturbance taken as an augmented/extended state. The state feedback controller is used to regulate the tracking error between the real output and a reference signal for the physical plant [13]. It should be remembered that the ESO together with the nonlinear plant takes the form of chain-like integrators where any of the linear or nonlinear control design methods, like those presented in [14–19] can be applied for feedback stabilization and performance enhancement.

The rest of the paper is organized as follows: in Section 2, paper scope and contribution are summarized. Problem statement and preliminaries with some mathematical definitions are given in Section 3. The main results of the proposed NLESO with the stability studies are presented in section Section 4. Section 5 describes an enhanced Version of ADRC with the proposed NLESO as its central core. The simulations of the enhanced ADRC on the permanent magnet DC motor are shown in Section 6. conclusions and future work are mentioned in Section 7.

## 2. Paper Scope and Contribution

In this paper, a new class of nonlinear extended state observers(NLESO) are proposed to actively reject the generalized disturbance for a general uncertain nonlinear system according to the principle of estimation/cancelation. For this NLESO, a saturation-like nonlinear error function was suggested to attenuate the large observer error at the starting stage and/or at the time of a discontinuous disturbance is injected to the system, consequently alleviating the peaking phenomenon. From the stability analysis of the error dynamics of NLESO observer, the finite-time stability analysis based on Lyapunov principle and Self-Stable Region(SSR) were introduced an applied on NLESO.

## 3. Problem Statement and Preliminaries

Assume an $n$-dimensional SISO nonlinear system which is expressed by [20],

$$y^n(t) = f\big(t, x_1(t), x_2(t), \ldots, x_n(t)\big) + w(t) + bu(t) \tag{1}$$

which can be rewritten as a chain of integrals with nonlinear uncertainties appearing in the $n\text{-}th$ equation,

$$\begin{cases} \dot{x}_1(t) = x_2(t), \ x_1(0) = x_{10}, \\ \dot{x}_2(t) = x_3(t), \ x_2(0) = x_{20}, \\ \quad \vdots \\ \quad \dot{x}_n(t) = f\big(t, x_1(t), x_2(t), \ldots, x_n(t)\big) + w(t) + bu(t), \ x_n(0) = x_{n0}, \\ y(t) = x_1(t) \end{cases}$$
(2)

where $u(t) \in C(\mathbb{R}, \mathbb{R})$ is the control input, $y(t)$ the measured output, $f(\cdot) \in C(\mathbb{R}^n, \mathbb{R})$ an unknown system function, $w(t) \in C(\mathbb{R}, \mathbb{R})$ the uncertain exogenous disturbance, $x(t) =$ $\left(x_1(t), x_2(t), \ldots, x_n(t)\right)^T$ the state vector of the nonlinear system and $x(0) = (x_{10}, x_{20}, \ldots, x_{n0})$ the initial state, $L(t) = f\left(t, x_1(t), x_2(t), \ldots, x_n(t)\right) + w(t)$ is called *"generalized disturbance"* [21]. By adding the extended state $x_{n+1}(t) \overset{\text{def}}{=} L(t) = f(t, .) + w(t)$, the system (1) can be written as,

$$\begin{cases} \dot{x}_1(t) = x_2(t), \ x_1(0) = x_{10}, \\ \dot{x}_2(t) = x_3(t), \ x_2(0) = x_{20}, \\ \quad \vdots \\ \dot{x}_n(t) = x_{n+1}(t) + bu(t), \ x(t) = x_{n0} \\ \quad \dot{x}_{n+1}(t) = \Delta(t) = \dot{f}\left(t, x_1(t), x_2(t), \ldots, x_n(t)\right) + \dot{w}(t), \ x_{n+1}(0) = x_{n+1,0}, \\ y(t) = x_1(t) \end{cases}$$

(3)

The following Linear ESO (LESO) needs to be designed to estimate the states of the nonlinear system as well as the generalized disturbance $L(t)$ and is described as [22],

$$\begin{cases} \dot{\hat{x}}_1(t) = \hat{x}_2(t) + \beta_1(y(t) - \hat{x}_1(t)), \\ \dot{\hat{x}}_2(t) = \hat{x}_3(t) + \beta_2(y(t) - \hat{x}_1(t)), \\ \quad \vdots \\ \dot{\hat{x}}_\rho(t) = \hat{x}_{\rho+1}(t) + bu(t) + \beta_\rho\left(y(t) - \hat{x}_1(t)\right), \\ \dot{\hat{x}}_{\rho+1}(t) = \beta_{\rho+1}\left(y(t) - \hat{x}_1(t)\right), \end{cases}$$

(4)

where $\hat{x}_{n+1}(t) = \hat{L} = \hat{f}\left(t, x_1(t), x_2(t), \ldots, x_\rho(t)\right) + \hat{w}(t)$, $\beta_i$ is a constant observer gain to be tuned, $i \in \{1, 2, \ldots, \rho, \rho + 1\}$. With $\beta_i = \frac{a_i}{\varepsilon^i}$, where $a_i$, $i \in \{1, 2, \ldots, \rho, \rho + 1\}$ are pertinent constants, and $\varepsilon$ is the constant gain or the reciprocal of observer's bandwidth, $\rho$ is the relative degree of the nonlinear system [20]. The observer gain is directly proportional to the observer bandwidth. Selecting a bandwidth that is too large will lead to a drop in the estimation error within an acceptable bound [23]. Therefore, the observer bandwidth is chosen to be sufficiently larger than the disturbance frequency and smaller than the frequency of unmodelled dynamics [24]. On the other hand, the performance of the ESO will be deteriorated if the bandwidth of the ESO is selected as either too low or too high. The side effects of adopting a large value for bandwidth can be summarised as, the measurement noise causes a degradation on the output tracking and the control signal. Some unmodelled high frequencies dynamics may be activated beyond a certain frequency causing inconsistency in the closed-loop system. A dynamical system whose states approaches as accurate as possible the states of (3) through time and coincides with them as $t \to \infty$.

Motivated by the above reasons, the aim is to design a nonlinear ESO(NLESO) is required to be designed whose Functions are as follows:

1. Its states approach as accurate as possible the states of (3) through time and coincide with them as $t \to \infty$.
2. It reduces the peaking phenomenon.
3. It avoids large transient behaviors.
4. It guarantees fast-convergence and robustness concerning noise.

A general nonlinear extended state observer is given by,

$$\begin{cases} \dot{\hat{x}}_1(t) = \hat{x}_2(t) + g_1(y(t) - \hat{x}_1(t)), \\ \dot{\hat{x}}_2(t) = \hat{x}_3(t) + g_2(y(t) - \hat{x}_1(t)), \\ \quad \vdots \\ \dot{\hat{x}}_\rho(t) = \hat{x}_{\rho+1}(t) + bu(t) + g_\rho\left(y(t) - \hat{x}_1(t)\right), \\ \dot{\hat{x}}_{\rho+1}(t) = g_{\rho+1}\left(y(t) - \hat{x}_1(t)\right). \end{cases}$$

(5)

If the nonlinear functions $g_i \colon \mathbb{R} \to \mathbb{R}, i \in \{1, 2, \ldots, \rho + 1\}$ were selected appropriately, the state variables of the nonlinear system could track the state variables of the original system and generalized disturbance. A nonlinear function is the mathematical fitting of "*big error, small gain or small error, big gain*" [25]. This function is generally selected as a nonlinear combination power function and is given as [26,27]

$$fal(e, \alpha, \delta) = \begin{cases} \frac{e}{\delta^{1-\alpha}} & |e| \leq \delta \\ |e|^{\alpha} sgn(e) & |e| > \delta \end{cases} \tag{6}$$

where $\delta$ is a small number which is used to express the length of the linear part [28]. The $fal(\cdot)$ is a piecewise continuous, nonlinear, saturation, a monotonous increasing function. The following assumptions are made for the next sections [29,30].

**Assumption (A1).** The function $f$ and $w(t)$ is continuously differentiable for all $(t, x(t)) \in \mathbb{R} \times \mathbb{R}^n$.

$$|u| + |f| + |\dot{w}| + \left| \frac{\partial f}{\partial t} \right| + \left| \frac{\partial f}{\partial x_i} \right| \leq c_o + \sum_{j=1}^{n} c_j |x_j|^k$$

for some positive constants $c_j, j = 0, 1, \ldots, n$ and positive integer $k$.

∎

**Assumption (A2).** The generalized disturbance $L(t)$ is bounded and belongs to a known compact set $L \subset \mathbb{R}$, *i.e*,

$$\sup_t \ L(t) \ \leq \infty$$

∎

**Assumption (A3).** There is a positive constant $M$ such that $|\Delta(t)| \leq M$ for $t \geq 0$, where

$$\Delta(t) = L(t) = \dot{f}\big(t, x_1(t), x_2(t), \ldots, x_n(t)\big) + \dot{w}(t) = \frac{d}{dt}\big[f(t, x(t)) + w(t)\big] = \frac{\partial f(t, x(t))}{\partial t} + \sum_{i=1}^{n-1} x_{i+1}(t) \frac{\partial f(t, x(t))}{\partial x_i} + \big[f\big(t, x_1(t), x_2(t), \ldots, x_n(t)\big) + w(t) + bu(t)\big] \frac{\partial f(t, x(t), w(t))}{\partial x_n} + \dot{w}(t) \tag{7}$$

**Assumption (A4).** The solution $x_i$ of (2) satisfy $|x_i(t)| \leq B$ for some constant $B > 0$ for all $i = 1, 2, \ldots, n$, and $t \geq 0$.

∎

Next, the convergence analysis of LESO for uncertain nonlinear single-input single-output is considered.

**Definition 2** [31]: Assuming that $S$ is a region in the state space, which contains the origin. If it satisfies the condition that any system's trajectory, which remains in it after a particular time, will eventually converge to the origin, then $S$ is called the self-stable region (SSR) of the system. ∎

In the next, we will derive a formula based on Lyapunov function which describes the convergence of the proposed NLESO. The assumptions given below are related to finite-time stability analysis of both the ESO and the closed-loop system.

**Lemma 1** [32,33]: the system

$$\dot{e} = -k \ sgn(e) \ |e|^{\alpha} \tag{8}$$

is globally finite-time stable where $k > 0$, $\alpha \in (0, 1)$. For any initial value of $e(t)$ at $t = t_o$, *i.e.*, $e_o$, it is easily obtained that the solution trajectory of (8) will reach $e = 0$ in finite time $t_f = \frac{|e_o|^{1-\alpha}}{(1-\alpha)k}$.

**Theorem 1** [33, 34]: Consider the nonlinear system $\dot{x} = f(x)$ with $f(0) = 0$. Suppose there exists a continuous function $V: \mathcal{D} \to \mathbb{R}$ on an open neighborhood $\mathcal{D} \subseteq \mathbb{R}^n$ of the origin such that the following conditions hold,

1. $V(x)$ is positive definite.
2. $\dot{V}(x) + cV^\alpha \leq 0$.

Then the origin is finite-time stable, and the settling time $t_f$ depending on the initial conditions $x(0) = x_0$ is given as,

$$t_f \leq \frac{V(x_0)^{1-\alpha}}{c(1-\alpha)} \tag{9}$$

for all $x_0$ in some open neighborhood of the origin, where $c > 1$, $0 < \alpha < 1$.

## 4. Main Results

In this section, the proposed NLESO will be presented followed by the stability analysis of its finite-time convergence based on self-stable region technique. The conversion of the mismatched disturbance into matched one will also be introduced.

### 4.1 The Proposed NLESO

The proposed NLESO for the general uncertain nonlinear system of (2) is designed as,

$$\begin{cases} \dot{\hat{x}}_1(t) = \hat{x}_2(t) + \beta_1 g_1(\omega_0(y(t) - \hat{x}_1(t))), \\ \dot{\hat{x}}_2(t) = \hat{x}_3(t) + \beta_2 g_2(\omega_0(y(t) - \hat{x}_1(t))), \\ \qquad\qquad \vdots \\ \dot{\hat{x}}_\rho(t) = \hat{x}_{\rho+1}(t) + bu(t) + \beta_\rho g_\rho(\omega_0(y(t) - \hat{x}_1(t))), \\ \dot{\hat{x}}_{\rho+1}(t) = \beta_{\rho+1} g_{\rho+1}(\omega_0(y(t) - \hat{x}_1(t))) \end{cases}$$

(10)

where $\hat{x}_{n+1}(t) = \hat{L}$, $\beta_i = a_i \omega_0^{i-1}$, $\omega_0 > 0$ is the ESO bandwidth and $a_i, i = 1, 2, \ldots, \rho + 1$, are selected according to,

$$a_i = \frac{(\rho+1)!}{i!(\rho+1-i)!}$$

such that the characteristic equation

$$s^{\rho+1} + a_1 s^\rho + \cdots \ldots + a_\rho s + a_{\rho+1} = (s+1)^{\rho+1}$$

is Hurwitz, where $\rho$ is the relative degree of the nonlinear system. The nonlinear function $g_i: \mathbb{R} \to \mathbb{R}$ is designed as:

$$g_i(\omega_0 e) = c_i \left( K_\alpha |\omega_0 e|^\alpha sgn(\omega_0 e) + K_\beta |\omega_0 e|^\beta \cdot \omega_0 e \right) \tag{11}$$

where $K_\alpha, K_\beta, \alpha, c_i$ and $\beta$ are the positive design parameters, $c_i$'s are used to help further reduce peaking phenomenon, a coherent problem with LESO, they are chosen such that $c_1 > c_1 > \cdots .. > c_{\rho+1}$. It should be noted that the proposed NLESO estimates the states of the uncertain nonlinear system up to relative degree $\rho$ of the nonlinear system. For the chain of integrals characterized by (2) or (3), *the relative degree is* $\rho = n$. So, the NLESO will estimate up to *n-th* states of (2) in addition to the generalized disturbance defined by $x_{n+1}(t)$.

The proposed nonlinear function has a saturation-like profile which obeys the principle of "*small error, large gain, and large error, small gain*", it is an odd function in terms of the error *e*. It has the following additional features:

- $g\ (0) = 0$
- $g\ (e) = k(e).\,e$ , where $\ k_{min} \le k(l) < \infty.$

*4.2 Stability Analysis of the proposed NLESO*

The convergence of the proposed NLESO is studied based on how well it estimates the states of the uncertain nonlinear system and the generalized disturbance, **Theorem** 3 below shows the convergence analysis in terms of the estimation error dynamics and finite-time stability.

**Theorem** 2: Consider the nonlinear system of (2) and assumptions 1-4 are satisfied, then proposed NLESO described by (11) is globally asymptotically stable, it is finite-time convergent to (2) with $t_f > t_o$ such that $e_i = 0, i = 1\ ,2, \dots., n + 1$ for all $t > t_f$.

**Proof**: By introducing the augmented state $x_{n+1} = L$ into (2), we obtained (3). Moreover, set

$$\begin{cases} e_i(t) = x_i(t) - \hat{x}_i(t) & i \in \{1, 2, \dots n\} \\ e_{n+1}(t) = x_{n+1}(t) - \hat{x}_{n+1}(t) & i = n + 1 \end{cases}$$

(12)

It should be noted that in (12) that $x_{n+1}(t) - \hat{x}_{n+1}(t) = L - \hat{L}$, where $L$ and $\ \hat{L}$ are the generalized disturbances and the estimated generalized disturbance respectively. A direct computation shows that the estimation error dynamics of (13) satisfies:

$$\begin{cases} \dot{x}_1(t) - \dot{\hat{x}}_1(t) = x_2(t) - (\hat{x}_2(t) + \beta_1 g_1(\omega_0(y(t) - \hat{x}_1(t)))) \\ \dot{x}_2(t) - \dot{\hat{x}}_2(t) = x_3(t) - (\hat{x}_3(t) + \beta_2 g_2(\omega_0(y(t) - \hat{x}_1(t)))) \\ \quad\quad\vdots \\ \dot{x}_n(t) - \dot{\hat{x}}_n(t) = x_{n+1}(t) + bu(t) - (\hat{x}_{n+1}(t) + bu(t) + \beta_n g_n(\omega_0(y(t) - \hat{x}_1(t)))) \\ \dot{x}_{n+1}(t) - \dot{\hat{x}}_{n+1}(t) = \Delta(t) - \beta_{n+1} g_{n+1}\left(\omega_0(y(t) - \hat{x}_1(t))\right) \end{cases}$$

(13) Substituting (11) in (13), gives

$$\begin{cases} \dot{e}_1(t) = e_2(t) - \beta_1 c_1 K_\alpha |\omega_0 e_1(t)|^\alpha sgn(\omega_0 e_1(t)) - \beta_1 c_1 K_\beta |\omega_0 e_1(t)|^\beta . \omega_0 e_1(t) \\ \dot{e}_2(t) = e_3(t) - \beta_2 c_2 K_\alpha |\omega_0\ \ e_1(t)|^\alpha sgn(\omega_0 e_1(t)) - \beta_2 c_2 K_\beta |\omega_0 e_1(t)|^\beta . \omega_0 e_1(t) \\ \quad\vdots \\ \dot{e}_n(t) = e_{n+1}(t) - \beta_n c_n K_\alpha |\omega_0 e_1(t)|^\alpha sgn(\omega_0 e_1(t)) - \beta_n c_n K_\beta |\omega_0 e_1(t)|^\beta . \omega_0 e_1(t) \\ \dot{e}_{n+1}(t) = \Delta(t) - \beta_{n+1} c_{n+1} K_\alpha |\omega_0 e_1(t)|^\alpha sgn(\omega_0 e_1(t)) - \beta_{n+1} c_{n+1} K_\beta |\omega_0 e_1(t)|^\beta . \omega_0 e_1(t) \end{cases}$$

where $\Delta(t)$ is the derivative of the generalized disturbance $L\ $ and is given by (7). According to Lemma 1 and Theorems 5-7 of [35], and if assumption **A3** holds true, then the error dynamics of (14) are asymptotically stable, i.e., the error $e_1(t)$ in the first equation approaches zero, so do $e_2(t), e_3(t), \dots \dots, e_{n+1}(t)$ go to zero. Moreover, the error dynamics of (14) are finite-time stable, i.e., it converges to (2) within $\ \ t_f > t_o$ such that $e_i = 0,\ i = 0, 1, \dots., n + 1$ For all $t > t_f$. To prove it is time-finite convergent, we use the self-stable region (SSR) approach in [31] to accomplish this task, and we proceed as follows,

Firstly, for simplicity assume *n* = 1, then the error dynamics of the NLESO is

$$\begin{cases} \dot{e}_1(t) = e_2(t) - \beta_1 c_1 K_\alpha |\omega_0 e_1(t)|^\alpha sgn(\omega_0 e_1(t)) - \beta_1 c_1 K_\beta |\omega_0 e_1(t)|^\beta . \omega_0 e_1(t) \\ \dot{e}_2(t) = \Delta(t) - \beta_2 c_2 K_\alpha |\omega_0 e_1(t)|^\alpha sgn(\omega_0 e_1(t)) - \beta_2 c_2 K_\beta |\omega_0 e_1(t)|^\beta . \omega_0 e_1(t) \end{cases}$$

(15)

Define $m_2(e_1, e_2) = e_2 - \beta_1 c_1 K_\alpha |\omega_0 e_1|^\alpha sgn(e_1) - \beta_1 c_1 K_\beta |\omega_0 e_1|^\beta \omega_0 e_1 + k q_1(e_1)\ sgn(e_1)$, where $q_1(e_1(t))$ is positive definite continuous function, i.e., $\ q_1(0) = 0$ , $k > 1$. Also, define

$S = \{e_1, e_2 : |m_2(e_1, e_2)| \leq q_1(e_1)\}$, assume that there exists $(e_1, e_2) \in S$, $\forall \ t > T$. According to the structure of $S$

$$e_2 - \beta_1 c_1 K_\alpha |\omega_0 e_1|^\alpha sgn(\omega_0 e_1) - \beta_1 c_1 K_\beta |\omega_0 e_1|^\beta \omega_0 e_1 + kq_1(e_1) \ sgn(e_1) \leq q_1(e_1)$$

or

$$e_2 - \beta_1 c_1 K_\alpha |\omega_0 e_1|^\alpha sgn(\omega_0 e_1) - \beta_1 c_1 K_\beta |\omega_0 e_1|^\beta \omega_0 e_1 \leq q_1(e_1) - kq_1(e_1) \ sgn(e_1) \tag{16}$$

Choose a Lyapunov candidate function $V(e_1(t))$ as

$$V(e_1(t)) = \int_0^{e_1(t)} e_1(t) de_1(t)$$

Then,

$$\dot{V}(e_1(t)) = e_1 \frac{de_1(t)}{dt}$$

$$= e_1 \big( e_2(t) - \beta_1 c_1 K_\alpha |\omega_0 e_1(t)|^\alpha sgn(\omega_0 e_1(t)) - \beta_1 c_1 K_\beta |\omega_0 e_1(t)|^\beta . \omega_0 e_1(t) \big) \tag{17}$$

Sub (17) in (18), we get

$$\dot{V}(e_1(t)) \leq |e_1|(q_1(e_1) - kq_1(e_1) \ ) \leq -(k-1)|e_1| \ q_1(e_1) \tag{18}$$

From the positiveness definition of $q_1(e_1)$ and since $k > 1$, we conclude that $\dot{V}(e_1(t))$ is negative definite. Hence,

$$t \to \infty \Longrightarrow e_1(t) \to 0$$

and based on the structure of $S$

$$t \to \infty \Longrightarrow e_2(t) \to 0$$

Let $q_1(e_1) = k_1 \ |e_1(t)|^\alpha$, with $k_1 > 0$ and $0 < \alpha < 1$. With this choice of $q_1(e_1)$, $\dot{V}(e_1(t))$ becomes

$$\dot{V}(e_1(t)) \leq -r|e_1|^{1+\alpha}$$

From (18), $V = \frac{e_1^2}{2}$ or $e_1 = (2V)^{\frac{1}{2}}$, and $\dot{V}(e_1(t))$ gets its final form as

$$\dot{V}(e_1(t)) \leq -r(2V)^{\frac{1+\alpha}{2}} \tag{19}$$

With $r = (k-1) \ k_1$. According to **Theorem** 1, the error dynamics of (16) is finite-time convergent with

$$t_f \ \leq \frac{V(e_1(t_0))^{1-\tilde{\alpha}}}{c(1-\tilde{\alpha})}$$

where $\tilde{\alpha} = \frac{1+\alpha}{2}$. ∎

**Remark 1**: The proof of the first part of Theorem 3 is based on the Filippov sense, where any discontinuous differential equation $\dot{x} = v(x)$, $x \in \mathbb{R}^n$ and $v$ is a locally bounded measurable vector function, is replaced by an equivalent differential inclusion $\dot{x} \in V(x)$ (see[36] ). While the second part was proved by self-stable region approach defined in definition 1, see [37] and the references therein. ∎

The dynamics of the proposed NLESO given by (10) can be represented in terms of (14) as,

$$\begin{cases} \dot{\hat{x}}_1(t) = \hat{x}_2(t) + e_2(t) - \dot{e}_1(t), \\ \qquad \dot{\hat{x}}_i(t) = \hat{x}_{i+1}(t) + e_{i+1}(t) - \dot{e}_i(t) \qquad , i = 1,..,n-1 \\ \qquad \dot{\hat{x}}_n(t) = \hat{x}_{n+1}(t) + e_{n+1}(t) - \dot{e}_n(t) + u(t) \\ \dot{\hat{x}}_{n+1}(t) = \Delta(t) - \dot{e}_{n+1}(t) \end{cases}$$

(20)

The dynamics (20) tells us that the states of the NLESO suffer from observer error dynamics (14).

*4.3 Mismatched Disturbance and System of Integrals Chain*

The original ESO in [2], [38] assumes that the plant is expressed in the Integral Chain Form (ICF) satisfying the matched condition (*Brunovsky* form) [39]. Therefore, its applicability is restricted to systems which, directly or using a change of variable, can be expressed in the ICF. Performing such transformation is not always easy as it is mentioned in [2], [40], especially if the system has zero-dynamics. Furthermore, in certain nonlinear systems, the disturbances appear in the system in a different channel of the control input and hence does not satisfy the matching condition. Consequently, the standard manipulation of ADRC for this mismatched disturbance is no longer available. For example, assume the following uncertain nonlinear system,

$$\begin{cases} \dot{x}_1 = f_1(x_1,....,x_n) + b_1 d_1 \\ \quad \vdots \\ \dot{x}_i = f_i(x_1,....,x_n) + b_i d_i \\ \dot{x}_n = f_i(x_1,....,x_n) + b_n u + b_n d_n \\ y = x_1 \end{cases}$$

(21)

where $x_i = [x_1, x_2, ... , x_n]^T \in \mathbb{R}^n$, $u \in \mathbb{R}$, $y$ are the states of the system, the control input, and the system output, respectively. $f_i$, are smooth function and are differentiable and $f_i(\cdot) \neq 0$ for $i = 1, 2, ... , n-1$, $d_i(t) \in \mathbb{R}$ represents the external mismatched disturbance, $d_n$ is the matched disturbance.

Therefore, motivated by the successful results of the ESO, it was recently pointed out in [41] that it is imperative to develop ESO-based control techniques for systems without assuming the ICF and satisfying the matching condition( the disturbance must appear on the same channel of the control input, i.e., *Brunovsky* form). The next theorem is proposed to deal with mismatch disturbances and uncertainties assuming $n = 2$ for simplicity.

**Theorem 4:** Consider the 2nd order affine nonlinear dynamical system with mismatched disturbance satisfying assumption **A2** represented by:

$$\begin{cases} \dot{x}_1 = f_1(x_1, x_2) + b_1 d \\ \dot{x}_2 = f_2(x_1, x_2) + b_2 u \\ y = x_1 \end{cases}$$

(22)

The above system can be transformed into the nonlinear model satisfying the matching condition (*Brunovsky* form) with the state-space given by:

$$\begin{cases} \dot{\tilde{x}}_1 = \tilde{x}_2 \\ \quad \dot{\tilde{x}}_2 = \hat{f}(\tilde{x}_1, \tilde{x}_2, x_2) + \hat{b}(u + \hat{d}) \\ y = \tilde{x}_1 \end{cases}$$

(23)

where  $\hat{f}(x_1, x_2) = \frac{\partial f_1(x_1, x_2)}{\partial x_1} f_1(x_1, x_2) + \frac{\partial f_1(x_1, x_2)}{\partial x_2} f_2(x_1, x_2),$

$$\hat{b} = b_2 \frac{\partial f_1(x_1, x_2)}{\partial x_2},$$

$$\hat{d} = (b_1 \frac{\partial f_1(x_1, x_2)}{\partial x_1} d + b_1 \dot{d}) / (b_2 \frac{\partial f_1(x_1, x_2)}{\partial x_2})$$

**Proof:** differentiate the first equation of (23) w.r.t $t$, one gets:

$$\ddot{x}_1 = \frac{\partial f_1(x_1, x_2)}{\partial x_1} \dot{x}_1 + \frac{\partial f_1(x_1, x_2)}{\partial x_2} \dot{x}_2 + b_1 \dot{d} \qquad (24)$$

Substitute (22) into (24) to get

$$\ddot{x}_1 = \frac{\partial f_1(x_1, x_2)}{\partial x_1} (f_1(x_1, x_2) + b_1 d) + \frac{\partial f_1(x_1, x_2)}{\partial x_2} (f_2(x_1, x_2) + b_2 u) + b_1 \dot{d} \qquad (25)$$

Rearrange (25), then,

$$\ddot{x}_1 = \frac{\partial f_1(x_1, x_2)}{\partial x_1} f_1(x_1, x_2) + \frac{\partial f_1(x_1, x_2)}{\partial x_2} f_2(x_1, x_2) + b_2 \frac{\partial f_1(x_1, x_2)}{\partial x_2} \left( u + \left( \frac{b_1 \dot{d} + b_1 \frac{\partial f_1(x_1, x_2)}{\partial x_1} d}{b_2 \frac{\partial f_1(x_1, x_2)}{\partial x_2}} \right) \right) \qquad (26)$$

Then (26) is reduced to

$$\ddot{x}_1 = \hat{f}(x_1, x_2) + \hat{b}(u + \hat{d})$$

Let,  $\tilde{x}_1 = x_1$   $\tilde{x}_2 = \dot{x}_1$. Then

$$\begin{cases} \dot{\tilde{x}}_1 = \tilde{x}_2 \\ \dot{\tilde{x}}_2 = \hat{f}(\tilde{x}_1, \tilde{x}_2, x_2) + \hat{b}(u + \hat{d}) \\ y = \tilde{x}_1 \end{cases}$$
$$(27)$$

What remained is just $x_2$, and one can find an expression of $x_2$ from the first equation of (22) and substitute this expression in (27) to get a matched nonlinear state-space equation in terms of the new coordinate system $\tilde{x}_1, \tilde{x}_2$. Finally, (27) is called the canonical form of ADRC ( *Brunovsky* form) [1].    ∎

To illustrate the aforementioned transformation, let apply it on the numerical example given by [42],

$$\begin{cases} \dot{x}_1 = x_2 + e^{x_1} + d \\ \dot{x}_2 = -2x_1 - x_2 + u \\ y = x_1 \end{cases}$$
$$(28)$$

It is clear that  $f_1(x_1, x_2) = x_2 + e^{x_1}, \ b_1 = 1, \ f_2(x_1, x_2) = -2x_1 - x_2, \ b_2 = 1.$ Let

$$\tilde{x}_1 = x_1 \qquad (29)$$

$$\tilde{x}_2 = f_1(x_1, x_2) + d = x_2 + e^{x_1} + d \qquad (30)$$

Then

$$\dot{\tilde{x}}_1 = \tilde{x}_2 \qquad (31)$$

$$\dot{\tilde{x}}_2 = \dot{\tilde{x}}_1 = \dot{x}_2 + e^{x_1} \dot{x}_1 + \dot{d} \qquad (32)$$

Sub. (28) in (32) to have:

$$\dot{\tilde{x}}_2 = -2x_1 - x_2 + u + e^{x_1}(x_2 + e^{x_1} + d) + \dot{d} \tag{33}$$

and sub. (29) in (33), Rearrange (34), results in

$$\dot{\tilde{x}}_2 = -2\tilde{x}_1 - x_2 + u + e^{\tilde{x}_1}x_2 + e^{2\tilde{x}_1} + e^{\tilde{x}_1}d + \dot{d} \tag{34}$$

Finally,

$$\hat{f}(\tilde{x}_1, \tilde{x}_2) = -2\tilde{x}_1 - x_2 + e^{2\tilde{x}_1} + x_2 \ e^{\tilde{x}_1} \ , \ \ \hat{d} = e^{\tilde{x}_1}d + \dot{d}.$$

Alternatively,

$$\hat{b} = b_2 \frac{\partial f_1(x_1, x_2)}{\partial x_2} = 1 \ \ .1 = 1$$

$$\hat{d} = (b_1 \frac{\partial f_1(x_1, x_2)}{\partial x_1}d + b_1\dot{d})/(b_2 \frac{\partial f_1(x_1, x_2)}{\partial x_2}) = \frac{1.e^{x_1}d + \dot{d}}{1} = 1.e^{x_1}d + \dot{d}$$

$$\hat{f}(x_1, x_2) = \frac{\partial f_1(x_1, x_2)}{\partial x_1}f_1(x_1, x_2) + \frac{\partial f_1(x_1, x_2)}{\partial x_2}f_2(x_1, x_2)$$
$$= e^{x_1}(x_2 + e^{x_1}) + 1.(\ -2x_1 - x_2)$$

Let $\ \ \tilde{x}_1 = x_1, \ \tilde{x}_2 = \ \dot{\tilde{x}}_1$, then

$$\begin{cases} \dot{\tilde{x}}_1 = \tilde{x}_2 \\ \quad \dot{\tilde{x}}_2 = \hat{f}(\tilde{x}_1, \tilde{x}_2, x_2) + \hat{b}(u + \hat{d}) \\ y = \tilde{x}_1 \end{cases}$$
$$(35)$$

With $\hat{f}(\tilde{x}_1, \tilde{x}_2, x_2) = x_2 e^{\tilde{x}_1} + e^{2\tilde{x}_1} - 2\tilde{x}_1 - x_2 \ , \ \ \hat{b} = 1 \ , \ \hat{d} = e^{x_1}d + \dot{d}$. One can further eliminate $x_2$ from (35) by substituting $x_2 = \tilde{x}_2 - e^{\tilde{x}_1} - d \ $ in (35) to get

$$\dot{\tilde{x}}_2 = -2\tilde{x}_1 - (\tilde{x}_2 - e^{\tilde{x}_1} - d) + u + e^{\tilde{x}_1}(\tilde{x}_2 - e^{\tilde{x}_1} - d) + e^{2\tilde{x}_1} + e^{\tilde{x}_1}d + \dot{d}$$

$$= -2\tilde{x}_1 - \tilde{x}_2 + e^{\tilde{x}_1} + d + u + \tilde{x}_2 e^{\tilde{x}_1} - e^{2\tilde{x}_1} - d + e^{2\tilde{x}_1} + e^{\tilde{x}_1}d + \dot{d}$$

$$= -2\tilde{x}_1 - \tilde{x}_2 + e^{\tilde{x}_1} + u + \tilde{x}_2 e^{\tilde{x}_1} + e^{\tilde{x}_1}d + \dot{d}$$

Which with the same $\ \hat{b} = 1, \ $ leads to,

$$\hat{f}_{new}(\tilde{x}_1, \tilde{x}_2) = -2\tilde{x}_1 - \tilde{x}_2 + e^{\tilde{x}_1} + \tilde{x}_2 e^{\tilde{x}_1},$$

$$\hat{d}_{new} = e^{\tilde{x}_1}d + \dot{d}$$

<div style="text-align: right;">∎</div>

It must be noted that the matched disturbance $\hat{d}$ or $\hat{d}_{new}$ is different from the original mismatched disturbance $d$ of (22) in the sense that after being transformed into the same channel of the control input, it is expressed in terms of the dynamic states of the nonlinear system and derivative of the original mismatched disturbance. In effect, the proposed NLESO will in real-time manner estimate and cancel $\hat{f}(\tilde{x}_1, \tilde{x}_2, x_2) + \hat{d}$ and depending on how well the NLESO estimate the dynamic states of the nonlinear system and the generalized disturbance, the nonlinear system together with the NLESO will look like a chain of integrals up to the relative degree $\rho$ of the original uncertain nonlinear system.

## 5. Application of The Proposed NLESO in ADRC

The classical Active Disturbance Rejection Control ( ADRC)  proposed by J. Han [2] is built by combining the tracking differentiator (TD), the nonlinear state error combination (NLSEF), and the linear extended state observer (LESO). In this work, an enhanced version of the ADRC(EADRC) which is called EADRC-NLESO is illustrated to emphasize that the proposed NLESO is employed in the design. It consists of a Second Order Nonlinear Differentiator (SOND) [43], an Improved Nonlinear State Error Feedback (INLSEF) controller [44], and the proposed NLESO. In the Improved INLSEF controller, the algorithm uses the $sign(.)$ together and the exponential function which are integrated as follows, $u_{INLSEF} = u_o = \Psi(e) = k(e)^T f(e) + u_{int}$, Where $e \in \mathbb{R}^n$  is  the vector of the state error, defined as $e = [e^{(0)} \quad .... \, e^{(i)} \, .... \quad e^{(n-1)}]^T$. In this regard, $e^{(i)}$ is the $i$-$th$ derivative of the state error defined as, $e^{(i)} = x^{(i)} - z^{(i)}$. The function  $k(e)$ and the function $f(e)$ are defined in [44]. It must be mentioned that $u_{int} = 0$ in our work, where integral action in ADRC is almost achieved by the ESO. Supposedly, the ESO will estimate and cancel online all the errors caused by any kind of discrepancy in the nonlinear system including the external disturbances.  The NLESO (for $n$ = 2) has the following state-space representation,

$$\begin{cases} \dot{\hat{x}}_1 = \hat{x}_2(t) + \beta_1 c_1[K_\alpha|\omega_0(y - \hat{x}_1(t))|^\alpha sign(\omega_0(y - \hat{x}_1(t))) + \\ \qquad K_\beta|\omega_0(y - \hat{x}_1(t))|^\beta(\omega_0(y - \hat{x}_1(t)))] \\ \dot{\hat{x}}_2 = \hat{x}_3(t) + bu + \beta_2 c_2[K_\alpha|\omega_0(y - \hat{x}_1(t))|^\alpha sign(\omega_0(y - \hat{x}_1(t))) + \\ \qquad K_\beta|\omega_0(y - \hat{x}_1(t))|^\beta(\omega_0(y - \hat{x}_1(t)))] \\ \dot{\hat{x}}_3 = \beta_3 c_3[K_\alpha|\omega_0(y - \hat{x}_1(t))|^\alpha sign(\omega_0(y - \hat{x}_1(t))) + \\ \qquad K_\beta|\omega_0(y - \hat{x}_1(t))|^\beta(\omega_0(y - \hat{x}_1(t)))] \end{cases} \qquad (36)$$

where $\hat{x}$  $(t) = [\hat{x}_1(t), \hat{x}_2(t), \hat{x}_3(t)]^T \in \mathbb{R}^3$, is a vector that includes the predictable states of the plant and the total-disturbance. The coefficients $\beta_1 = 3$, $\beta_2 = 3\omega_o$, $\beta_3 = \omega_o^2$, $K_\alpha$,  $\alpha, K_\beta$, $c_1$, $c_2$ , $c_3$ and $\beta \in \mathbb{R}^+$ are NLESO design parameters.

Another structure of ADRC was designed to compare the performance of the proposed EADRC-NLESO with it. It has the same configuration of the EADRC-NLESO, but with LESO instead of NLESO, throughout the simulations, it is referred to as EADRC-LESO. Based on the above, the control signal which actuates the nonlinear system in the ADRC paradigm is given by

$$v = u_o - \frac{\hat{x}_3(t)}{b} \qquad (37)$$

## 6. Simulations Results

As an application of the EADRC, the following numerical simulations include the nonlinear control of Permanent Magnet DC (PMDC) motor shown in Figure 1 with Coulomb friction force using EADRC.  The nonlinear model of the PMDC motor including the external disturbance is of mismatched type, see (25). Applying Newton's law and Kirchoff's law, we get the following equations,

$$\begin{cases} J_{eq}\frac{d^2\theta}{dt^2} = T - T_L - b_{eq}\frac{d\theta}{dt} \\ L\frac{di}{dt} = -Ri + v - e \end{cases}$$

$$= \begin{cases} \frac{d^2\theta}{dt^2} = \frac{1}{J_{eq}}(K_t i - T_L - b_{eq}\frac{d\theta}{dt}) \\ \frac{di}{dt} = \frac{1}{L}(-Ri + v - K_b\frac{d\theta}{dt}) \end{cases}$$

(38)

where $v$ is the input voltage applied to the motor (Volt) , $K_b$ is electromotive force constant constant (Volt / rad/s), $K_t$ is the torque constant (N.m/A) , $R$ is the electric resistance constant(Ohm), $L$ is the electric self-inductance (Henry), $J_{eq}$ is the total-equivalent moment of inertia(kg.m²), $J_{eq} = J_m + J_L/N^2$, where $J_L$ is the load moment of inertia (kg.m²), $J_m$ is the motor armature moment of inertia(kg.m²), $b_{eq}$ is the total-equivalent viscous damping of the combined motor rotor and load (N.m/rad.s), $b_{eq} = b_m + b_L/N^2$, $b_m$ is the motor's rotor damping (N.m/rad.s), and $b_L$ is the load viscous damping (N.m/rad.s), $N$ is the gearbox ratio, $T_L$ (N.m) is the load torque applied at the motor side.

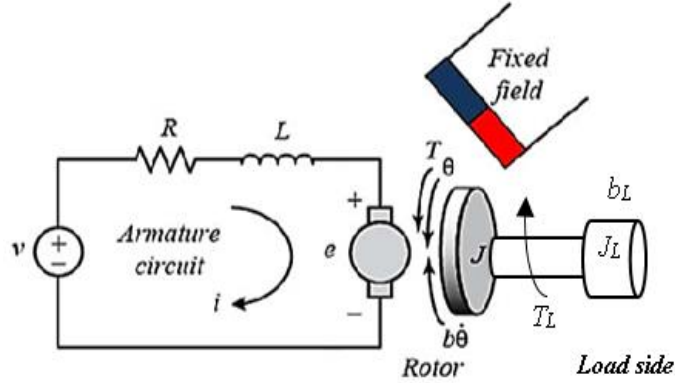

**Figure** 1. Schematic diagram of the PMDC motor.

Applying the transformation of appendix A, we get a simplified model for the nonlinear state-space representation of the PMDC motor expressed in Brunovsky form given by [3],

$$\begin{cases} \dot{x}_1 = x_2 \\ \dot{x}_2 = -\frac{R \ b_{eq} + K_t K_b}{L \ J_{eq}} x_1 - \frac{(L \ b_{eq} + R \ J_{eq})}{L \ J_{eq}} x_2 + \frac{K_t}{L \ J_{eq}}(v + d) \\ y = x_1/N \end{cases}$$

(39)

As can be seen, *the relative degree $\rho = n$*. The state $x_1$ is the angular velocity (rad/s) of the PMDC motor before dividing the actual angular velocity of the motor shaft $\frac{d\theta}{dt}$ by the gear ratio $N$, while the output $y$ is measured after the gearbox, *i.e.*, $y = \frac{1}{N}x_1$. The state $x_2$ is the angular acceleration (rad/s²). The equivalent disturbance at the input $d = -\frac{L}{K_t}\dot{T}_L - \frac{R}{K_t}T_L$ and $T_L$ (N.m) is the load torque applied at the motor side. The load torque is given as

$$T_L = \frac{1}{N}(T_{ext} + F_c sgn(x_1))$$

(40)

where $F_c$ is the Coulomb friction force [45], it is nonlinear function of $x_1$, that is why the system of (26) is nonlinear, $T_{ext}$ is the external disturbance torque, usually of discontinuous type. The values of the parameters for PMDC motor are [46]: $R_a = 0.1557$, $L_a = 0.82$, $K_b = 1.185$, $K_t = 1.1882$, $n = 3.0$, $J_{eq} = 0.2752$, and $b_{eq} = 0.3922$, $F_c = 1$. The parameters of the proposed EADRC-NLESO are as follows, INLSEF: $k_{11} = 1.95599$, $k_{12} = 1.22208$, $k_{21} = 0.50231$, $k_{22} = 3.2652$, $\mu_1 = 4.92537$, $\mu_2 = 3.74434$, $\alpha_1 = 0.693947$, $\alpha_2 = 0.770208$. The SOND: $a = 0.97893$, $b = 5.58718$, $c = 8.38639$, $\sigma = 26.5$. The NLESO: $\omega_o = 35$, $K_{it} = 0.99927$, $\alpha = 0.301361$, $K_\beta = 0.38$, $\beta = 0.305151$, $\beta_1 = 3$, $\beta_2 = 105$, $\beta_3 = 1225$, $c_1 = 0.5$, $c_2 = 0.125$, $c_3 = 0.0625$. While the parameters of the EADRC-LESO are, INLSEF: $k_{11} = 1.76353$, $k_{12} = 0.719549$, $k_{21} = 0.762186$, $k_{22} = 3.04664$, $\mu_1 = 8.69763$, $\mu_2 = 2.35869$, $\alpha_1 = 0.688673$, $\alpha_2 = 0.644945$. The SOND: $a = 0.97893$, $b = 5.58718$, $c = 8.38639$, $\sigma = 26.5$. The LESO: $\omega_o = 35$, $\beta_1 = 105$, $\beta_2 = 3675$, $\beta_3 = 42875$.

The PMDC controlled by both EADRC-NLESO and EADRC-LESO is tested by applying a reference angular- velocity equals to 1 rad per second at $t = 0$ and for $T_f = 10$ sec. To verify "Peaking Phenomenon", the initial conditions of both NLESO and LESO were set to $\hat{x}_1(0) = 0.5$, $\hat{x}_2(0) = $ $\hat{x}_3(0) = 0$, and that of the PMDC motor were, $x_1(0) = x_2(0) = x_3(0) = 0$. The results are shown in Figures 2 and 3.

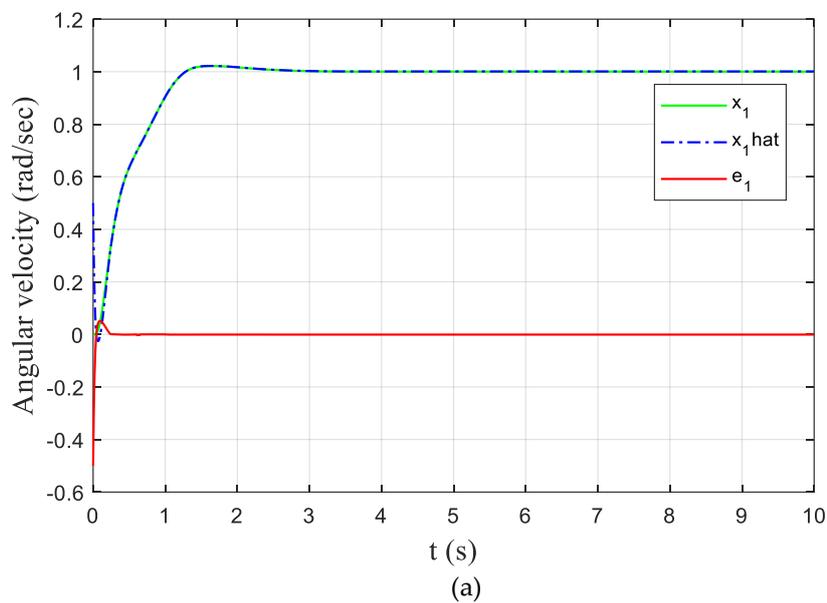

(a)

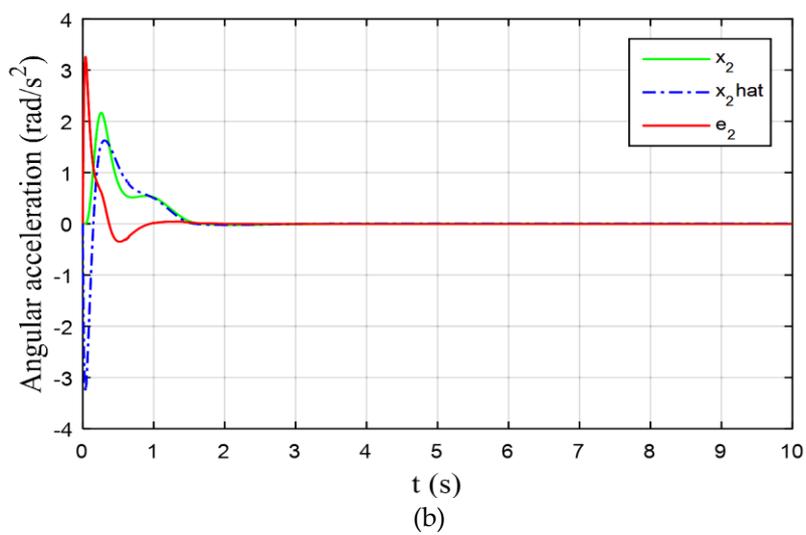

(b)

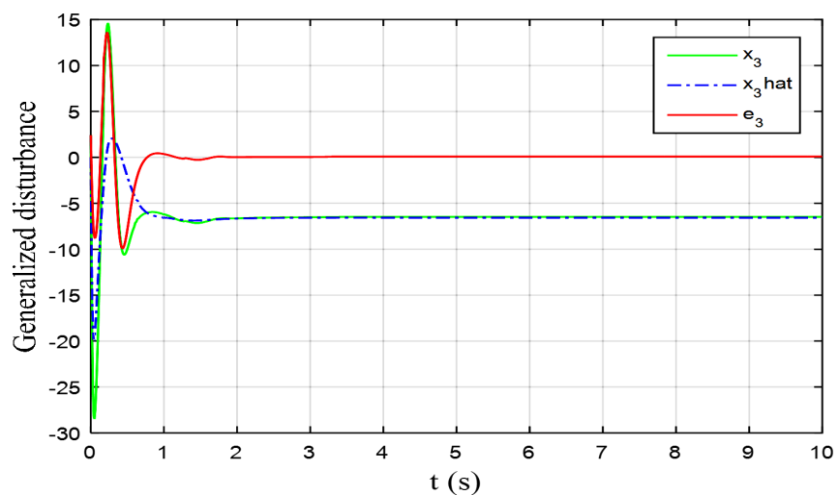

(c)

Fig. 2 Continued...

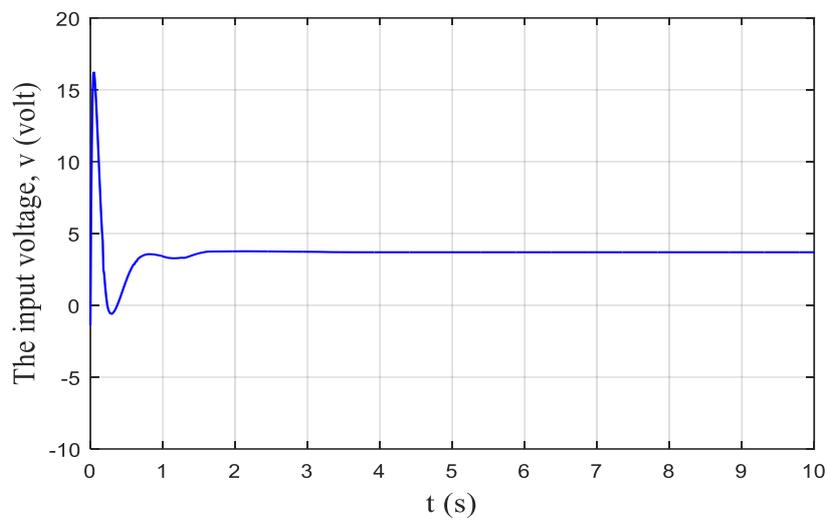

(d)

**Figure 2.** Results of the numerical simulations of the PMDC motor by EADRC-NLESO.

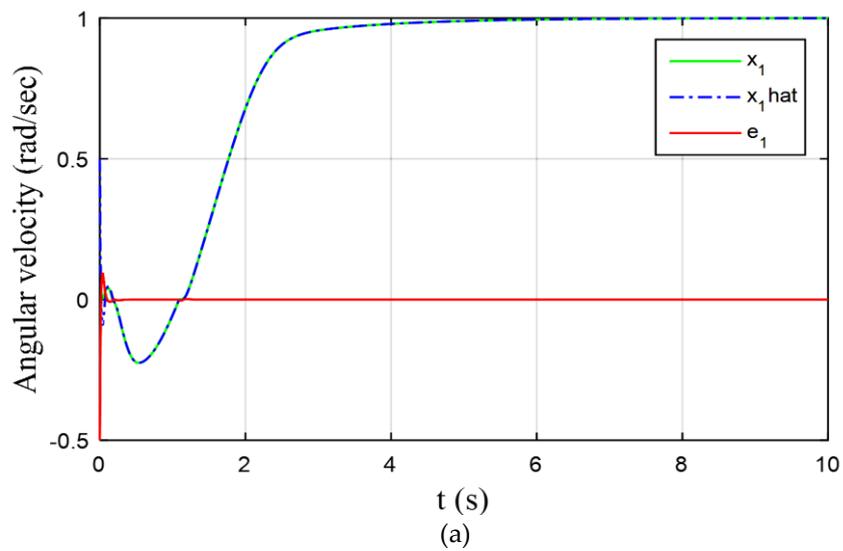

(a)

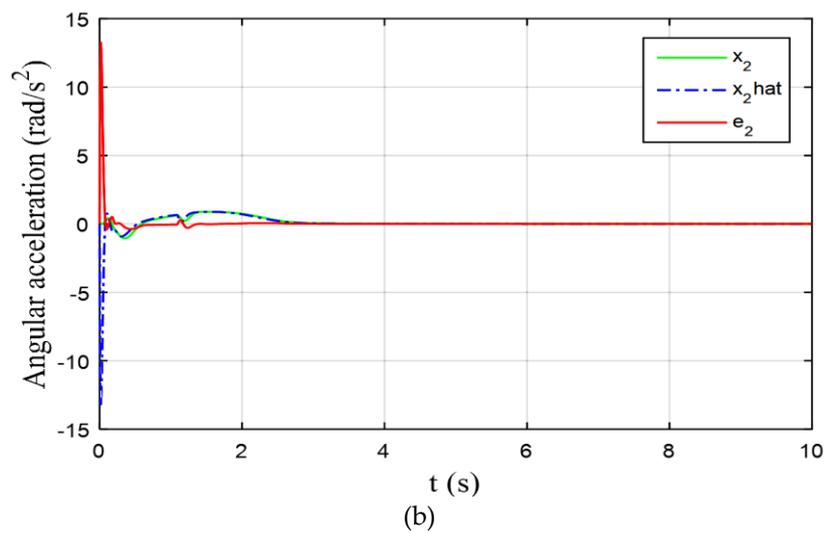

(b)

Fig. 3 Continued…

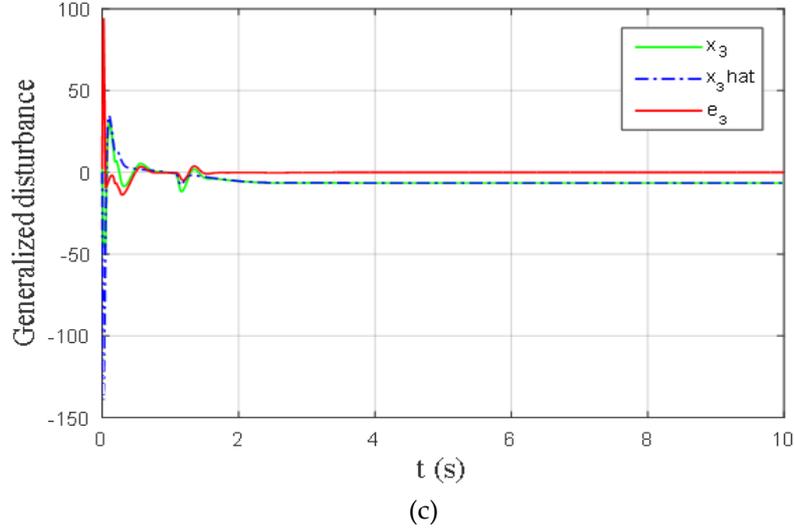

(c)

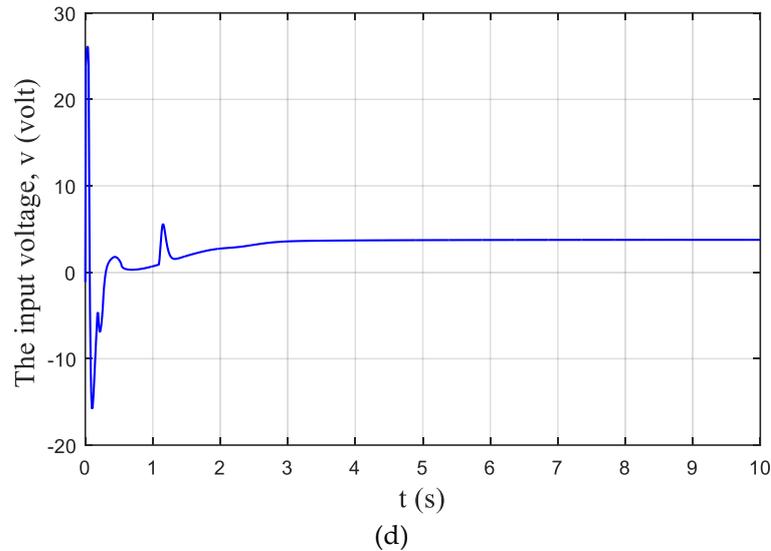

(d)

**Figure 3.** Results of the numerical simulations of the PMDC motor by EADRC-LESO.

For $\omega_o = 35$, the output response of the PMDC motor system of (40) using EADRC-NLESO are plotted in Figure 2. The angular velocity $x_1(t)$ and its estimation $\hat{x}_1(t)$ are drawn in Figure 2 (a), while the angular acceleration $x_2(t)$ and its estimation $\hat{x}_2(t)$ are plotted in Figure 2 (b). The generalized disturbance $x_3(t)$ together with its estimation are depicted in figures 2 (c). It can be seen that the estimation using NLESO is almost guaranteed. For the same $\omega_o$, the numerical results of the PMDC motor of (39) are redrawn in Figure 3, but using EADRC-LESO. It was clear that the LESO satisfactorily achieve state and generalized disturbance estimation, but suffers from "Peaking phenomenon", where $\hat{x}_1(t)$ peaks to -0.225, $\hat{x}_2(t)$ to -13.3 and $x_3(t)$ to -139.6, compared to NLESO where $\hat{x}_1(t)$ peaks to -0.026, $\hat{x}_2(t)$ to -3.27 and $x_3(t)$ did not peak. The under estimation in generalized disturbance $\hat{x}_3(t)$ (i.e. $x_3(t)$ does not exactly follow $x_3(t)$ ) can be treated successfully by increasing $\omega_o$, but on the account of noise filtration. The control signal in EADRC-LESO peaks down to -16 volts and up to 26 volts, whereas it peaks just to about 16.5 volts in EADRC-NLESO. It obvious that peaking in EADRC-NLESO is much smaller than that of EADRC -LESO.

To investigate the performance of the proposed NLESO to an exogenous disturbances, an experiment was conducted with an external torque acting as a step disturbance equal to 2 N.m ((2N.m /3)=0.666 N.m seen from the rotor side) is applied after the gearbox during the simulation at $t$ = 5 sec using MATLAB Simulink environment. The numerical results are shown in Figure 4. From this figure, it is easy to verify that both methods cancel the effect of the disturbance on the angular velocity efficiently with the EADRC-NLESO exhibits an undershoot larger than that of the EADRC-LESO (see Figure 3 (a), at $t$ = 5). The Integration-Time-Absolute-Error (ITAE) performance measure defined as,

$$\text{ITAE} = \int_0^{10} t \times |r - y| \ dt$$

where $y$ is the PMDC motor angular velocity output, and $r$ is the reference signal is used to measure the performance of both NLESO and LESO at steady-state. Its value in EADRC-LESO is 2.238968, and the ITAE value using EADRC-NLESO is 0.485433. It is clear that EADRC-NLESO outperforms EADRC-LESO significantly. Also, the control signal in the EADRC-LESO had a high peak at the starting and fluctuated after that until it reached the time of disturbance occurrence, again it overshoots to 9.3 volts. On the other hand, the control signal in the EADRC-NLESO overshoots with positive values only and to the half of that in EADRC-LESO. This leads to an increase of the energy required in the EADRC-LESO case, where an energy index defined as the integral square of the control signal ($u$) denoted as ISU is used to measure how much energy the control scheme requires, *i.e.*

$$\text{ISU} = \int_0^{10} \ u^2 dt$$

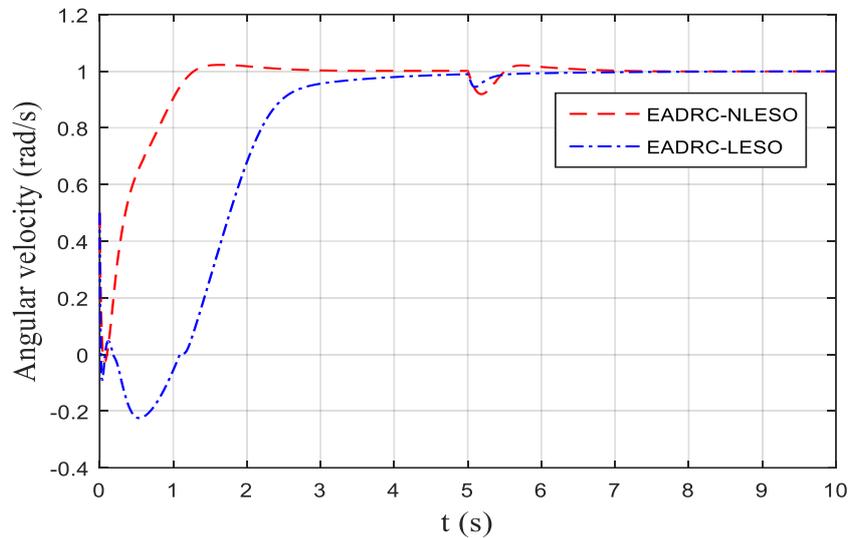

(a)

Figure 4. Continued...

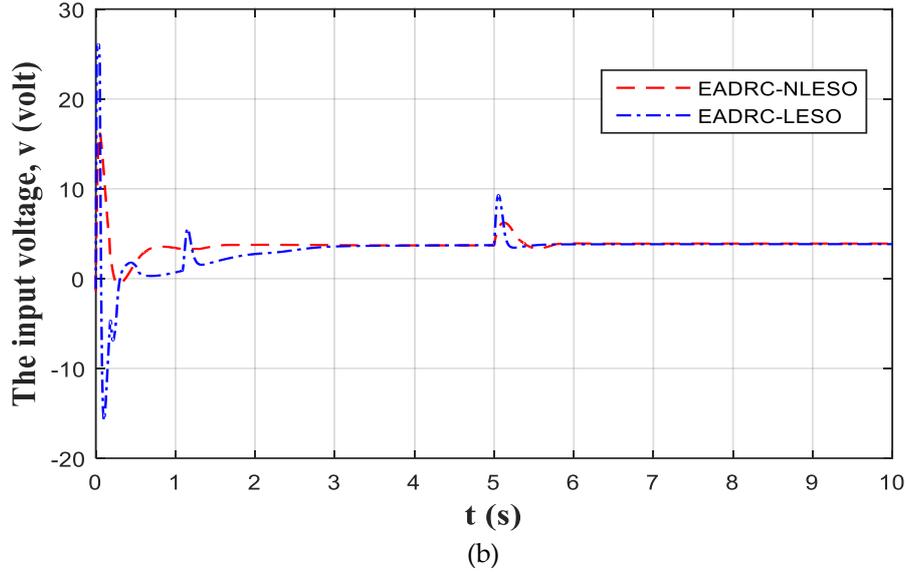

**Figure** 4. Angular velocity due to an external step disturbance of 2 N.m.

Based on ISU index, the control energy in EADRC-NLESO is 161.60068 and 172.92265 in EADRC-LESO. It must be mentioned that a limiter of $\mp12$ volts has been placed before the PMDC motor to limit the control signal input within the safe bounds.

Another experiment has been conducted to test the proposed NLESO against measurement noise. Assume that $y(t)$ has been contaminated by $n(t)$ usually (Gaussian) distributed random signal

$$y_n(t) = y(t) + n(t)$$

where $n(t)$ is usually (Gaussian) distributed random signal with $36 \times 10^{-6}$ and zero mean and is added using MATLAB Simulink block called *random number*. With the same values of the parameters in the above simulations including the bandwidth ($\omega_o$), the results are presented in Figure 5. It can be seen that the noise is still exist in the output response of the EADRC-LESO, while EADRC-NLESO produces a smoother response and suppresses noise evidently.

Finally, we end our simulations by subjecting the PMDC motor of (40) to parameter uncertainties and test the efficiency of EADRC-LESO, EADRC-NLESO, and compared to each other. Let the parameters to be varied are defined in their allowable range as follows,

$$J_{eq} = \bar{J}_{eq}(1 + \Delta_J \delta_J) \quad , \quad b_{eq} = \bar{b}_{eq}(1 + \Delta_b \delta_b) \quad , \quad R = \bar{R}(1 + \Delta_R \delta_R)$$

where $\bar{J}_{eq} = 0.2752$ , $\bar{b}_{eq} = 0.3922$, and $\bar{R} = 0.1557$ are called the nominal values of $J_{eq}$, $b_{eq}$, and $\Delta_J$, $\Delta_b$, and $\Delta_R$ are the possible relative changes in their respective parameters. Assume that $\delta_J = \delta_b = \delta_R = -1$ and let $\Delta_J = 0.2$, $\Delta_b = 0.4$, and $\delta_R = 0.5$, the angular velocity of the PMDC motor of (39) using both EADRC-LESO and EADRC-NLESO are graphed in Figure 6.

To end this section, let us discuss the advantages and disadvantages of using NLESO over LESO in ADRC configuration. At steady state, the error dynamics in (14) or (15) will be zero, i.e. $\dot{e}_1(t) = \dot{e}_2(t) = \dot{e}_3(t) = 0$, then , one can find a relationship between error $e_3(t)$ in the last equation of (15) in terms of $\Delta$, $\beta_3$, $c_3$, and other parameters, $e_3(t) = \varphi(\Delta, \beta_3, c_3, ..)$, where $\varphi(.)$ is a nonlinear function. The same can be done for $e_2(t)$ and $e_1(t)$, i.e $e_2(t) = \gamma(\Delta, \beta_2, ....)$, $e_3(t) = \vartheta(\Delta, \beta_1, ....)$, and this makes the errors of the NLESO are sensitive to $\Delta$, which is the rate the generalized disturbance $L(t)$, while in LESO, the errors are linearly depending on $\Delta$. That explains the jumps in the generalized disturbance using EADRC-NLESO at the points in the times where $L(t)$ exhibits a sudden increase or decrease (e.g., sudden external disturbance) or parameter variations in the nonlinear system of the PMDC motor described in (39). Figures 4(a) and 6 clarify this reasoning.

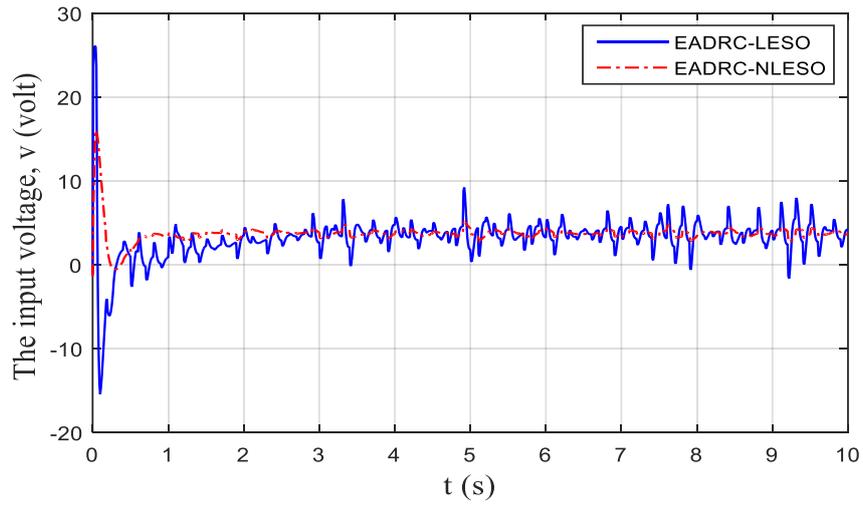

(a)

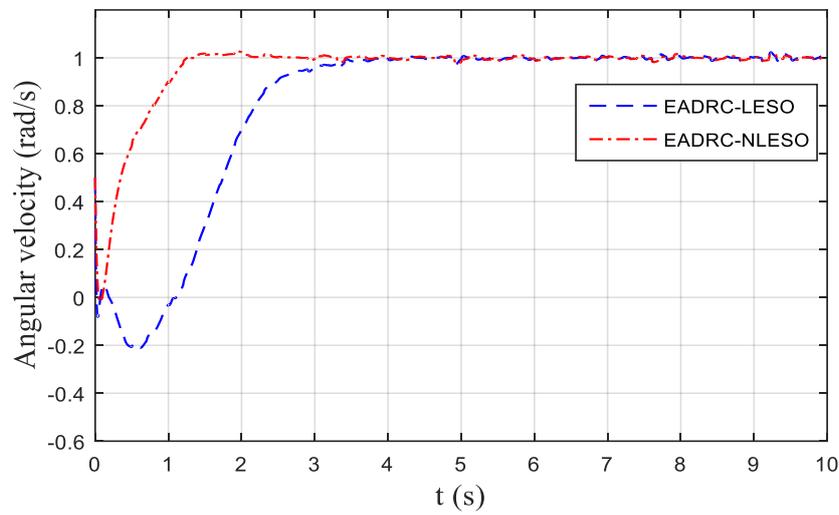

(b)

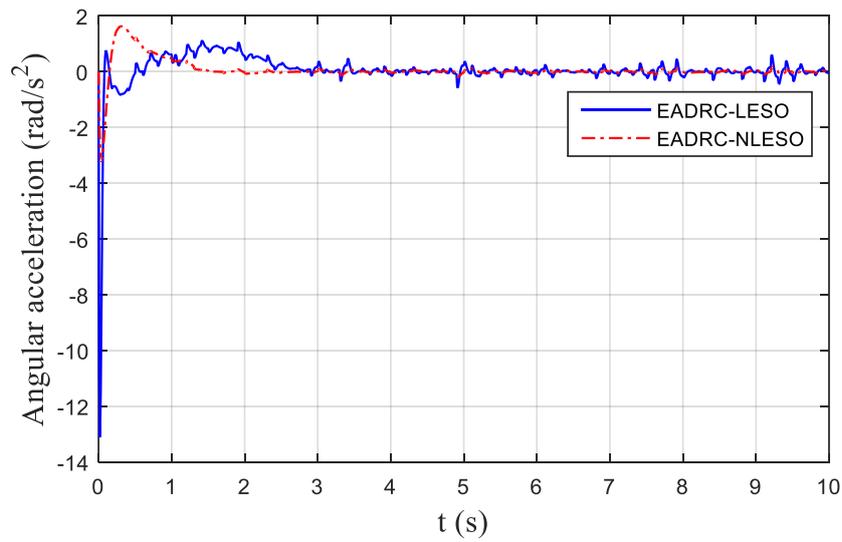

(c)

**Figure** 5. Continued…

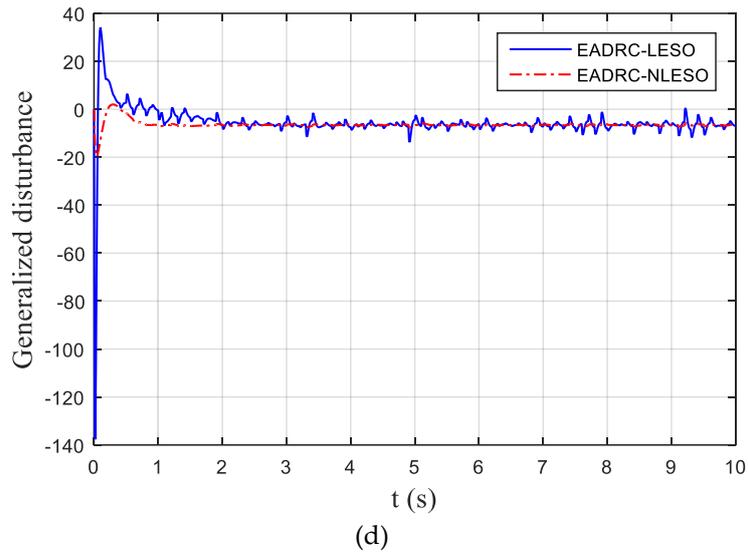

(d)

**Figure** 5. The numerical results for PMDC motor of (39) with measurement noise.

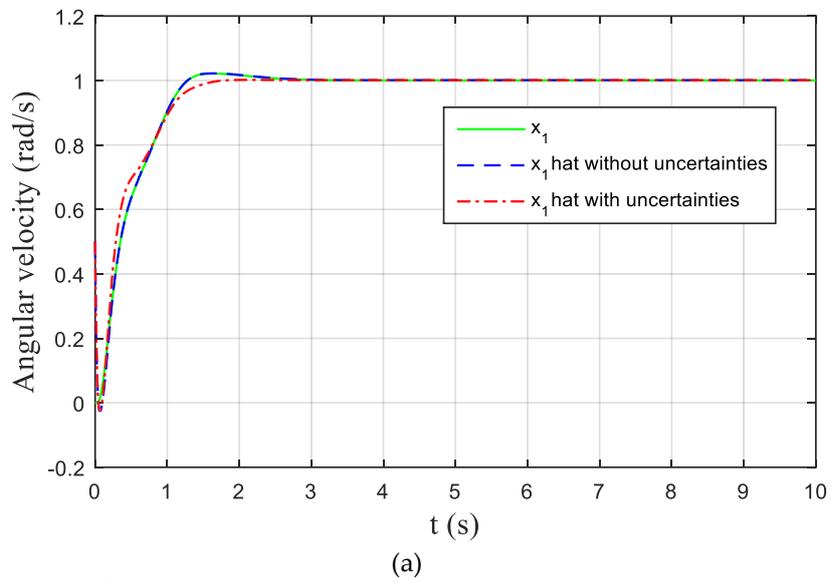

(a)

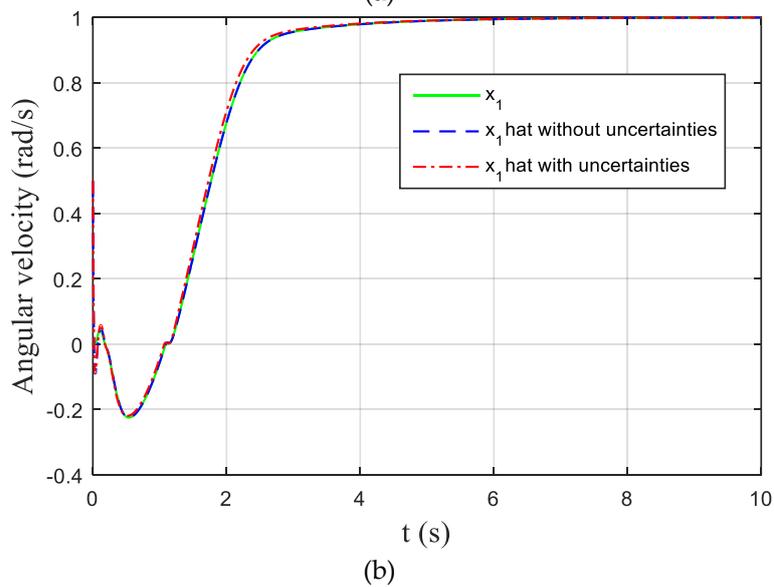

(b)

Figure 6. The angular velocity of PMDC motor of (39) with parameter uncertainties.

Next, we comment on the peaking phenomenon in both EADRC-LESO and EADRC-NLESO scenarios; it occurs at the starting when there is an initial condition $\hat{x}_1(0)$ for $\hat{x}_1(t)$ different from that of $x_1(t)$ ($i.e.\,\hat{x}_1(0) \neq x_1(0)$), this makes the terms in the equations of (14) that depends on $\omega_o$, $i.e.$ $a_i\omega_0^{i-1}\ g_i(\omega_0(y(t) - \hat{x}_1(t)))$ very large for some $g_i$'s and with large $\omega_0$. This term with a large value in the right hand side makes the ESO to produce large fluctuation on its output channels. For example when $g_i(e) = e$, that means the ESO acts as a LESO, then the term $a_i\omega_0^{i-1}\ g_i(\omega_0(y(0) - \hat{x}_1(0)))$ will have a high value for $y(0) - \hat{x}_1(0)$ and large $\omega_0$. The nonlinear function of (12) in our proposed NLESO (11) has a saturation-like behaviour for large $e_i$, in addition to the attenuating factors $c_i(c_1 > c_1 > \cdots.. > c_{n+1})$, that explains why our proposed NLESO has a smaller peaking than LESO for the same $\omega_0$. for example the $\omega_0$-dependent term in the $i$-$th$ equation of (14) can be expressed as

$$a_i\omega_0^{i-1}c_iK_\alpha\left|\omega_0(y(0) - \hat{x}_1(0))\right|^\alpha + a_i\omega_0^{i-1}c_iK_\beta\left|\omega_0(y(0) - \hat{x}_1(0))\right|^\beta. \omega_0. e_1(0)$$

The $K_\beta$ has small value in most cases as compared to $K_\alpha$, so for the easiness of illustration, we neglect the second term and the $\omega_0$-dependent becomes $a_ic_iK_\alpha\ \omega_0^{\alpha+i-1}\left|(y(0) - \hat{x}_1(0))\right|^\alpha$, and since $\alpha + i - 1 < i$, then $\omega_0^{\alpha+i-1} < \omega_0^i$ in the LESO for big $\omega_0$. For example assume $\omega_0 = 35$, $\alpha = 0.3$, $n = 2$, and $\left|(y(0) - \hat{x}_1(0))\right| = 1$. In the third equation ($i = 3$), the $\omega_0$-dependent term is $1.\frac{1}{16}$ . $0.99927.(35)^{0.3+3-1}.1 = 222.4521$, whereas the corresponding term in the LESO is $(\omega_0)^3 = (35)^3 = 42875$. The reduction in the peaking with the proposed saturation-like function $g_i(.)$ defined in (11) in our proposed NLESO of (11) is noteworthy .

The same reasoning can be extended to illustrate why the noise is attenuated using our proposed NLESO of (11) with the saturation-like function $g_i(.)$ defined in (11), where the noise is magnified to $(\omega_0)^3. n(t)$ in the $i$-$th$ equation of the LESO of (4). While with our proposed saturation-like function $g_i$ (.) defined in (11), the magnification of the noise $n(t)$ in our proposed NLESO is $a_ic_iK_\alpha\ \omega_0^{\alpha+i-1}\left|(y(0) - \hat{x}_1(0))\right|^\alpha$ which is proven to be less than $(\omega_0)^3$. The results of Figure 5 illustrate this justification.

# 7. Conclusions

In this paper, a novel saturation-like function has been proposed and employed in the design of a nonlinear extended state observer used to estimate the states and the generalized disturbance of any uncertain nonlinear system with mismatch disturbance. Stability analysis based on Lyapunov principles has shown the asymptotic convergence of the proposed ESO and finite-time stability is always guaranteed provided that the generalized disturbance is bounded. The advantage of the proposed ESO is that it produced smaller peaking and had immunity against measurement noise and parameter variations. All the mathematical investigations and the conducted experiments included in this paper proved that the proposed ESO presented better performance than the linear ESO for the mentioned reasons. Employing the proposed ESO in the ADRC configuration provided an excellent tool to control any uncertain nonlinear system and to counteract the generalized disturbance. As a future direction, this work can be extended to a general MIMO uncertain nonlinear system and apply the proposed ESO to non-affine control system like ball-and-beam system.

**Author Contributions**: Investigation, A. J. H; Supervision, I. K. I.

# Appendix A

*Conversion of Nonlinear PMDC motor mismatched model into a Brunovsky form*

Let $x_1 = \frac{d\theta}{dt}$ , $x_2 = i$ , then, the mismatched nonlinear mathematical model of the PMDC motor can written as

$$\begin{cases} \dot{x}_1 = \frac{1}{J_{eq}}(K_t x_2 - T_L - b_{eq} x_1) \\ \dot{x}_2 = \frac{1}{L}(-R x_2 + v - K_b x_1) \end{cases}$$

(A.1)

Let

$$\tilde{x}_1 = x_1,$$

$$\tilde{x}_2 = \dot{x}_1 = \frac{1}{J_{eq}}(K_t x_2 - T_L - b_{eq} x_1) \tag{A.2}$$

Then,

$$\dot{\tilde{x}}_1 = \tilde{x}_2$$

$$\dot{\tilde{x}}_2 = \frac{1}{J_{eq}}(K_t \dot{x}_2 - \dot{T}_L - b_{eq} \dot{x}_1) \tag{A.3}$$

Sub. (A.1) into (A.3), we get

$$\dot{\tilde{x}}_2 = \frac{1}{J_{eq}}[K_t \frac{1}{L}(-R x_2 + v - K_b \tilde{x}_1) - \dot{T}_L - b_{eq}\frac{1}{J_{eq}}(K_t x_2 - T_L - b_{eq}\tilde{x}_1)]$$

$$= -x_2\left[\frac{K_t R}{J_{eq}L} + \frac{b_{eq}K_t}{J_{eq}^2}\right] + \tilde{x}_1\left[\frac{b_{eq}^2}{J_{eq}^2} - \frac{K_t K_b}{J_{eq}L}\right] + \frac{K_t}{J_{eq}L}v + \frac{b_{eq}}{J_{eq}^2}T_L - \frac{1}{J_{eq}}\dot{T}_L \tag{A.4}$$

To express (A.4) in the new coordinate system ( $\tilde{x}_1$, $\tilde{x}_2$), we need to eliminate $x_2$ from (A.4). From (A.2),

$$x_2 = \frac{1}{K_t}(J_{eq}\tilde{x}_2 + T_L + b_{eq}\tilde{x}_1) \tag{A.5}$$

Sub. (A.5) in (A.4) to have

$$\dot{\tilde{x}}_2 = -\frac{1}{K_t}(J_{eq}\tilde{x}_2 + T_L + b_{eq}\tilde{x}_1)\left[\frac{K_t R}{J_{eq}L} + \frac{b_{eq}K_t}{J_{eq}^2}\right] + \tilde{x}_1\left[\frac{b_{eq}^2}{J_{eq}^2} - \frac{K_t K_b}{J_{eq}L}\right] + \frac{K_t}{J_{eq}L}v + \frac{b_{eq}}{J_{eq}^2}T_L - \frac{1}{J_{eq}}\dot{T}_L$$

$$= -\frac{R}{L}\tilde{x}_2 - \frac{R}{J_{eq}L}T_L - \frac{b_{eq}R}{J_{eq}L}\tilde{x}_1 - \frac{b_{eq}}{J_{eq}}\tilde{x}_2 - \frac{b_{eq}}{J_{eq}^2}T_L - \frac{b_{eq}^2}{J_{eq}^2}\tilde{x}_1 + \frac{b_{eq}^2}{J_{eq}^2}\tilde{x}_1 - \frac{K_t K_b}{J_{eq}L}\tilde{x}_1 + \frac{K_t}{J_{eq}L}v + \frac{b_{eq}}{J_{eq}^2}T_L - \frac{1}{J_{eq}}\dot{T}_L$$

$$= -\frac{R}{L}\tilde{x}_2 - \frac{R}{J_{eq}L}T_L - \frac{b_{eq}R}{J_{eq}L}\tilde{x}_1 - \frac{b_{eq}}{J_{eq}}\tilde{x}_2 - \frac{K_t K_b}{J_{eq}L}\tilde{x}_1 + \frac{K_t}{J_{eq}L}v - \frac{1}{J_{eq}}\dot{T}_L$$

$$= -\left(\frac{R}{L} + \frac{b_{eq}}{J_{eq}}\right)\tilde{x}_2 - \left(\frac{b_{eq}R}{J_{eq}L} + \frac{K_t K_b}{J_{eq}L}\right)\tilde{x}_1 + \frac{K_t}{J_{eq}L}(v + d)$$

where $d = -\frac{R}{K_t}T_L - \frac{L}{K_t}\dot{T}_L$ . So,

$$\begin{cases} \dot{\tilde{x}}_1 = \tilde{x}_2 \\ \dot{\tilde{x}}_2 = -\left(\frac{R}{L} + \frac{b_{eq}}{J_{eq}}\right)\tilde{x}_2 - \left(\frac{b_{eq}R}{J_{eq}L} + \frac{K_t K_b}{J_{eq}L}\right)\tilde{x}_1 + \frac{K_t}{J_{eq}L}(v + d) \end{cases}$$

(A.6)

which is exactly the model given by (39) and the load torque is given as $T_L = T_{ext} + F_c sgn(x_1)$.